\documentclass[a4paper,11pt]{article}
\pdfoutput=1

\usepackage[a4paper,vmargin={30mm,30mm},hmargin={30mm,30mm}]{geometry}

\usepackage{lmodern}
\usepackage[T1]{fontenc}
\usepackage[utf8]{inputenc}

\usepackage[inline]{enumitem}

\usepackage{amsfonts}
\usepackage{amstext}
\usepackage{amsmath}
\usepackage{mathtools}

\usepackage{arydshln}
\usepackage[margin=\parindent,labelfont=bf]{caption}

\usepackage{pbox}

\usepackage[usenames,dvipsnames]{xcolor}

\usepackage{tikz}
\usetikzlibrary{decorations.markings}
\usetikzlibrary{arrows}

\numberwithin{equation}{section}
\numberwithin{figure}{section}

\usepackage[
pdfauthor={
  Nils Kanning, 
  Yumi Ko, 
  Matthias Staudacher
}, 
pdftitle={
  Grassmannian Integrals as Matrix Models for Non-Compact Yangian Invariants
},
colorlinks=true,
linkcolor=RedViolet,
citecolor=Red,
urlcolor=Magenta,
unicode=true,
hypertexnames=false
]{hyperref}

\usepackage{cite}
\usepackage[parsep]{collref}

\DeclareMathOperator{\D}{d\!}
\DeclareMathOperator{\tr}{tr}
\DeclareMathOperator{\Min}{min}

\usepackage[yyyymmdd]{datetime}
\usepackage{currfile}

\usepackage[backgroundcolor=red!10!white,bordercolor=red,linecolor=red]{todonotes}

\begin{document} 

\baselineskip 5mm

\begin{titlepage}

  \hfill
  \pbox{5cm}{
    \texttt{HU-Mathematik-2014-38}\\
    \texttt{HU-EP-14/63}\\
    \texttt{AEI-2014-065}
  }
  
  \vspace{2\baselineskip}

  \begin{center}

    \textbf{\LARGE Graßmannian Integrals as Matrix Models\\[1ex]
    for Non-Compact Yangian Invariants}

    \vspace{2\baselineskip}

    Nils Kanning\textsuperscript{\textit{a,b}},
    Yumi Ko\textsuperscript{\textit{a}} and
    Matthias Staudacher\textsuperscript{\textit{a}}
 
    \vspace{2\baselineskip}

    \textit{
      \textsuperscript{a} 
      Institut für Mathematik und Institut für Physik, Humboldt-Universität zu Berlin,\\
      IRIS-Adlershof, Zum Großen Windkanal 6, 12489 Berlin, Germany\\
      \vspace{0.5\baselineskip}
      \textsuperscript{b}
      Max-Planck-Institut für Gravitationsphysik, Albert-Einstein-Institut,\\
      Am Mühlenberg 1, 14476 Potsdam, Germany\\
      \vspace{0.5\baselineskip}
    }

    \vspace{2\baselineskip}

    \texttt{
      \href{mailto:kanning@mathematik.hu-berlin.de}{kanning@mathematik.hu-berlin.de},
      \href{mailto:koyumi@mathematik.hu-berlin.de}{koyumi@mathematik.hu-berlin.de},
      \href{mailto:matthias@mathematik.hu-berlin.de}{matthias@mathematik.hu-berlin.de}
    }

    \vspace{2\baselineskip}
    
    \textbf{Abstract}
    
  \end{center}

  \noindent
  In the past years, there have been tremendous advances in the field
  of planar $\mathcal{N}=4$ super Yang-Mills scattering amplitudes. At
  tree-level they were formulated as Graßmannian integrals and were
  shown to be invariant under the Yangian of the superconformal
  algebra $\mathfrak{psu}(2,2|4)$. Recently, Yangian invariant
  deformations of these integrals were introduced as a step towards
  regulated loop-amplitudes. However, in most cases it is still
  unclear how to evaluate these deformed integrals. In this work, we
  propose that changing variables to oscillator representations of
  $\mathfrak{psu}(2,2|4)$ turns the deformed Graßmannian integrals
  into certain matrix models. We exemplify our proposal by formulating
  Yangian invariants with oscillator representations of the
  non-compact algebra $\mathfrak{u}(p,q)$ as Graßmannian
  integrals. These generalize the Brezin-Gross-Witten and
  Leutwyler-Smilga matrix models. This approach might make elaborate
  matrix model technology available for the evaluation of Graßmannian
  integrals. Our invariants also include a matrix model formulation of
  the $\mathfrak{u}(p,q)$ R-matrix, which generates non-compact
  integrable spin chains.

\end{titlepage}

\section{Introduction}
\label{sec:introduction}

The maximally supersymmetric Yang-Mills theory in four-dimensions, for
short $\mathcal{N}=4$ SYM, is a remarkably rich mathematical
model. Even more so in the planar limit where the theory is
conjectured to be integrable. By now this integrability is well
established for the spectral problem of anomalous dimensions, see the
comprehensive review series \cite{Beisert:2010jr}. Less is known about
integrability for scattering amplitudes. However, at tree-level the
amplitudes can be encoded as surprisingly simple formulas, so-called
Graßmannian integrals \cite{ArkaniHamed:2009dn,Mason:2009qx}, see also
\cite{ArkaniHamed:2012nw}. The mere existence of such formulas already
hints at an underlying integrable structure. Furthermore, it was shown
that tree-level amplitudes are invariant under the Yangian of the
superconformal algebra $\mathfrak{psu}(2,2|4)$
\cite{Drummond:2009fd}. For the Graßmannian integral formulation this
was achieved in \cite{Drummond:2010qh,Drummond:2010uq}. The appearance
of this infinite-dimensional Yangian algebra is synonymous with
integrability. Later, it was observed that the tree-level amplitudes
allow for multi-parameter deformations while maintaining Yangian
invariance. These deformations are of considerable interest as they
relate the four-dimensional scattering problem to the two-dimensional
quantum inverse scattering method. Furthermore, they might regulate
infrared divergences at loop-level \cite{Ferro:2012xw,Ferro:2013dga}.

As in the undeformed case, the deformed tree-level amplitudes can be
nicely packaged as Graßmannian integrals
\cite{Ferro:2014gca,Bargheer:2014mxa}. Let us briefly review this
formulation. The \emph{Graßmannian} $\text{Gr}(N,K)$ is the space of
all $K$-dimensional linear subspaces of $\mathbb{C}^N$. The entries of
a $K\times N$ matrix $C$ provide ``homogeneous'' coordinates on this
space. The transformation $C\mapsto VC$ with $V\in GL(K)$ corresponds
to a change of basis within a given subspace, and thus it does not
change the point in the Graßmannian. This allows us to describe a
generic point in $\text{Gr}(N,K)$ by the ``gauge fixed'' matrix
\begin{align}
  \label{eq:grassm-matrix}
  C=
  \left(
  \begin{array}{c:c}
    1_{K\times K}&\mathcal{C}\\
  \end{array}
  \right)
  \quad
  \text{with}
  \quad
  \mathcal{C}=
  \begin{pmatrix}
    C_{1 K+1}&\cdots&C_{1 N}\\
    \vdots&&\vdots\\
    C_{K K+1}&\cdots&C_{K N}\\
  \end{pmatrix}.
\end{align}
The amplitudes are labeled by the number of particles $N$ and the
degree of helicity violation $K$. Amplitudes with $K=2$ are maximally
helicity violating ($\text{MHV}$). The deformed $N$-point
$\text{N}^{K-2}\text{MHV}$ tree-level amplitude is given by the
\emph{Graßmannian integral}
\begin{align}
  \label{eq:grass_amp}
  \mathcal{A}_{N,K}=\int \D \mathcal{C}
  \frac{\delta^{4K|4K}(C\mathcal{W})}
  {(1,\ldots,K)^{1+v_{K}^+-v_1^-}\cdots (N,\ldots,K-1)^{1+v_{K-1}^+-v_N^-}}\,
\end{align}
with the holomorphic $K(N-K)$-form $\D\mathcal{C}=\bigwedge_{k,l}\D
C_{k l}$. In this formula $(i,\ldots,i+K-1)$ denotes the minor of the
matrix $C$ consisting of the consecutive columns
$i,\ldots,i+K-1$. These are counted modulo $N$ such that they are in
the range $1,\ldots,N$.  The kinematics of the $j$-th particle is
encoded in a supertwistor with components $\mathcal{W}^j_{A}$, where
$A$ is a fundamental $\mathfrak{gl}(4|4)$ index. The $2N$ deformation
parameters $\{v_i^+,v_i^-\}$ have to obey the constraints
\begin{align}
  \label{eq:spec-perm}
  v^+_{i+K}=v^-_i
\end{align}
for $i=1,\ldots,N$. Then the Graßmannian integral \eqref{eq:grass_amp}
is invariant under the Yangian of $\mathfrak{psu}(2,2|4)$, where the
generators of the algebra act on the supertwistors. In the undeformed
case $v_i^{\pm}=0$, the proper integration contour for
\eqref{eq:grass_amp} is known and the integral can be evaluated by
means of a multi-dimensional residue theorem
\cite{ArkaniHamed:2009dn,ArkaniHamed:2012nw}. In the deformed case,
the evaluation is much more involved due to branch cuts of the
integrand. Most notably there are partial results on the $6$-point
$\text{N}\text{MHV}$ amplitude \cite{Ferro:2014gca}. However, finding
an appropriate multi-dimensional integration contour for the
evaluation of \eqref{eq:grass_amp} is still a pressing open problem.

In the present work, we establish a connection between the Graßmannian
integral formulation of Yangian invariants and unitary matrix
models. We follow a systematic approach and do not focus on the
particular supertwistor realization of the algebra
$\mathfrak{psu}(2,2|4)$ that is often employed for
amplitudes. Instead, we work with a class of harmonic oscillator
representations of the non-compact algebra $\mathfrak{u}(p,q)$, where
we restrict to the bosonic case for clarity. We find that also in this
setting Yangian invariants can be formulated as Graßmannian
integrals. The \emph{only} change compared to \eqref{eq:grass_amp} is
that the delta function of the supertwistors gets replaced by an
exponential function of oscillators,
\begin{align}
  \label{eq:replace}
  \delta^{4K|4K}(C\mathcal{W}) \mapsto
  (\det\mathcal{C})^{-q} 
  e^{\tr(\mathcal{C}\mathbf{I}_\bullet^t+\mathbf{I}_\circ \mathcal{C}^{-1})}|0\rangle\,.
\end{align}
For the moment, we restrict for simplicity to the ``split helicity''
case $N=2K$ in order for $\mathcal{C}^{-1}$ to exist.  The $K\times K$
matrices $\mathbf{I}_\bullet$ and $\mathbf{I}_\circ$ contain certain
oscillator invariants associated with the compact subalgebras
$\mathfrak{u}(p)$ and $\mathfrak{u}(q)$, respectively. The integration
contour is still unspecified. We observe that the deformation
parameters $v_i^\pm$ can be chosen such that the exponents of all
minors in \eqref{eq:grass_amp} vanish. If we restrict in addition the
range of integration to unitary matrices $\mathcal{C}$, the
Graßmannian integral reduces to an intensively studied unitary matrix
model, the Brezin-Gross-Witten model
\cite{Gross:1980he,Brezin:1980rk}. Similarly, we may also obtain the
Leutwyler-Smilga model \cite{Leutwyler:1992yt}. This motivates us to
conjecture that the ``unitary contour'' works as well for general
deformation parameters. This would mean that the Graßmannian integrals
can be considered as novel types of unitary matrix models.  We provide
a non-trivial example of this conjecture by investigating the
invariant with $(N,K)=(4,2)$. In this example the Graßmannian integral
becomes a $U(2)$ matrix model that correctly evaluates to the
$\mathfrak{u}(p,q)$ R-matrix, which is known to be Yangian
invariant. This R-matrix generates non-compact integrable spin chains.

The connection between Graßmannian integrals and matrix models opens
exciting possibilities. In particular, advanced matrix model
technology such as character expansions, see e.g.\ the concise review
\cite{Morozov:2009jv}, might become applicable for the evaluation of
Graßmannian integrals. We expect our results to generalize
straightforwardly from $\mathfrak{u}(p,q)$ to superalgebras
$\mathfrak{u}(p,q|r)$ and thus to $\mathfrak{psu}(2,2|4)$. Hence our
matrix model approach should also be of utility for the open problem
mentioned above, the evaluation of deformed $\mathcal{N}=4$ SYM
amplitudes. There are further fascinating prospects which we elaborate
on in the outlook of section~\ref{sec:conclusions}.

\section{Yangian and Non-Compact Oscillators}
\label{sec:yangian}

In this preparatory section, we introduce the Yangian of the Lie
algebra $\mathfrak{gl}(n)$ and the notion of Yangian invariants. In
addition, we define the classes of oscillator representations of the
algebra $\mathfrak{u}(p,q)\subset\mathfrak{gl}(p+q=n)$ that we will
use to build up representations of the Yangian.

The \emph{Yangian} of $\mathfrak{gl}(n)$ is defined by the relation,
see e.g.\ \cite{Molev:2007},
\begin{align}
  \label{eq:def-rmm}
  R(u-u')(M(u)\otimes 1)(1\otimes M(u'))
  =(1\otimes M(u'))(M(u)\otimes 1)R(u-u')\,.
\end{align}
Here $R(u)$ acts on the tensor product $\mathbb{C}^n\otimes
\mathbb{C}^n$ and solves the Yang-Baxter equation. It is built from
$n\times n$ matrices with components
$(e_{AB})_{CD}=\delta_{AC}\delta_{DB}$ and reads
\begin{align}
  \label{eq:r-matrix}
  R(u)=1+u^{-1}\sum_{A,B=1}^ne_{AB}\otimes e_{BA}\,.
\end{align}
The operator valued monodromy matrix $M(u)$ contains the infinitely
many Yangian generators $M_{AB}^{(r)}$ with $r=1,2,\ldots$\,. They are
obtained from an expansion in the complex spectral parameter $u$
\begin{align}
  \label{eq:yangian-gen}
  M(u)=\sum_{A,B=1}^ne_{AB}M_{AB}(u)\,,\quad
  M_{AB}(u)=M_{AB}^{(0)}+u^{-1}M_{AB}^{(1)}+u^{-2}M_{AB}^{(2)}+\ldots
\end{align}
with $M_{AB}^{(0)}=\delta_{AB}$. Expressed in terms of these
generators, the defining relation \eqref{eq:def-rmm} becomes
\begin{align}
  \label{eq:def-gen}
  [M_{AB}^{(r)},M_{CD}^{(s)}]
  =\sum_{q=1}^{\Min(r,s)}
  \Big(M_{CB}^{(r+s-q)}M_{AD}^{(q-1)}-M_{CB}^{(q-1)}M_{AD}^{(r+s-q)}\Big)\,.
\end{align}
From this formula one easily deduces that all generators
$M_{AB}^{(r)}$ with $r>2$ can be expressed via $M_{AB}^{(1)}$ and
$M_{AB}^{(2)}$. In our study we are interested in states that are
\emph{Yangian invariant} \cite{Frassek:2013xza}
\begin{align}
  \label{eq:def-yi}
  M_{AB}(u)|\Psi\rangle=\delta_{AB}|\Psi\rangle\,.
\end{align}
With the help of the expansion in \eqref{eq:yangian-gen} this
condition translates into
\begin{align}
  \label{eq:def-yi-gen}
  M_{AB}^{(1)}|\Psi\rangle=0\,,\quad M_{AB}^{(2)}|\Psi\rangle=0\,.
\end{align}
From now on we specialize on realizations of the Yangian where the
monodromy is that of an inhomogeneous spin chain with $N$ sites. Thus
\begin{align}
  \label{eq:mono}
  M(u)=L_1(u-v_1)\cdots L_N(u-v_N)
\end{align}
is the product of $N$ Lax operators
\begin{align}
  \label{eq:lax}
  L_i(u-v_i)=1+(u-v_i)^{-1}\sum_{A,B=1}^ne_{AB}J^i_{BA}\,.
\end{align}
Here the meaning of the word \emph{inhomogeneous} is twofold. First,
we associate a complex inhomogeneity parameter $v_i$ with each
site. Second, each site carries a different representation
of the $\mathfrak{gl}(n)$ algebra with generators $J_{AB}^i$ that
satisfy
\begin{align}
  \label{eq:gln-alg}
  [J_{AB}^i,J_{CD}^i]=\delta_{CB}J_{AD}^i-\delta_{AD}J_{CB}^i
\end{align}
and act on a space $\mathcal{V}^i$. Consequently the matrix elements
of the monodromy $M(u)$ act on the tensor product
$\mathcal{V}^1\otimes\cdots\otimes \mathcal{V}^N$. The Yangian
generators introduced in \eqref{eq:yangian-gen} can be expressed in
terms of the $\mathfrak{gl}(n)$ generators,
\begin{align}
  \label{eq:yangian-gen-explicit}
  M^{(1)}_{AB}=\sum_{i=1}^N J_{BA}^i\,,\quad
  M^{(2)}_{AB}=\sum_{i=1}^Nv_i J_{BA}^i
  +\sum_{\substack{i,j=1\\i<j}}^N\sum_{C=1}^n J_{CA}^i J_{BC}^j\,,\quad
  \ldots\,.
\end{align}

Next, we introduce the representations of the $\mathfrak{gl}(n)$
algebra which we will employ at the sites of the spin chain monodromy
\eqref{eq:mono}. We work with certain classes of unitary
representations of the \emph{non-compact} Lie algebra
$\mathfrak{u}(p,q)\subset \mathfrak{gl}(p+q)$ that are constructed in
terms of a single family of \emph{harmonic oscillator algebras}. These
are sometimes referred to as ``ladder representations'', see e.g.\
\cite{Todorov:1966zz}. Consider the family of oscillator algebras
\begin{align}
  \label{eq:osc}
  [\mathbf{a}_A,\bar{\mathbf{a}}_B]=\delta_{AB}\,,\quad 
  \mathbf{a}_A^\dagger=\bar{\mathbf{a}}_A\,,\quad
  \mathbf{a}_A|0\rangle=0\,
\end{align}
with $A,B=1,2,\ldots n$. These oscillators are realized on a Fock
space $\mathcal{F}$ that is spanned by monomials of creation operators
$\bar{\mathbf{a}}_A$ acting on the vacuum $|0\rangle$. We split the
index $A=(\alpha,\dot{\alpha})$ into a pair of indices
$\alpha=1,\ldots,p$ and $\dot{\alpha}=p+1,\ldots,p+q$ with
$p+q=n$. Employing this notation, we define generators
\begin{align}
  \label{eq:osc-sym-gen}
  (\mathbf{J}_{AB})=
  \left(
    \begin{array}{c:c}
    \mathbf{J}_{\alpha\beta}&\mathbf{J}_{\alpha\dot\beta}\\[0.3em]
    \hdashline\\[-1.0em]
    \mathbf{J}_{\dot\alpha\beta}&\mathbf{J}_{\dot\alpha\dot\beta}\\
    \end{array}
  \right)
  =
  \left(
    \begin{array}{c:c}
    \bar{\mathbf{a}}_\alpha\mathbf{a}_\beta&
    -\bar{\mathbf{a}}_\alpha\bar{\mathbf{a}}_{\dot\beta}\\[0.3em]
    \hdashline\\[-1.0em]
    \mathbf{a}_{\dot\alpha}\mathbf{a}_\beta&
    -\mathbf{a}_{\dot\alpha}\bar{\mathbf{a}}_{\dot\beta}\\
    \end{array}
  \right)
\end{align}
that satisfy the $\mathfrak{gl}(n)$ algebra \eqref{eq:gln-alg}. Let
$\mathcal{V}_c\subset\mathcal{F}$ be the eigenspace of the central
element $\mathbf{C}=\sum_{A=1}^n\mathbf{J}_{AA}$ with eigenvalue
$c$. For each $c\in\mathbb{Z}$ this infinite-dimensional space forms a
unitary irreducible representation of $\mathfrak{u}(p,q)$. Hence we
may interpret $c$ as a representation label. The space $\mathcal{V}_c$
contains a lowest weight state, which by definition is annihilated by
all $\mathbf{J}_{AB}$ with $A>B$. Notice that in the special case
$q=0$ or $p=0$ the space $\mathcal{V}_c$ is finite-dimensional and
forms a unitary irreducible representation of the compact Lie algebra
$\mathfrak{u}(n)$. According to \eqref{eq:def-yi-gen}, Yangian
invariants are in particular $\mathfrak{gl}(n)$ singlet states. For
such states to exist, we need also spin chain sites with
representations that are dual to the class of representations
$\mathcal{V}_c$. Its generators are obtained from
\eqref{eq:osc-sym-gen} by
$\bar{\mathbf{J}}_{AB}=-\mathbf{J}_{AB}^\dagger$. This yields
\begin{align}
  \label{eq:osc-dual-gen}
  (\bar{\mathbf{J}}_{AB})=
  \left(
    \begin{array}{c:c}
    \bar{\mathbf{J}}_{\alpha\beta}&\bar{\mathbf{J}}_{\alpha\dot\beta}\\[0.3em]
    \hdashline\\[-1.0em]
    \bar{\mathbf{J}}_{\dot\alpha\beta}&\bar{\mathbf{J}}_{\dot\alpha\dot\beta}\\
    \end{array}
  \right)
  =
  \left(
    \begin{array}{c:c}
    -\bar{\mathbf{a}}_\beta\mathbf{a}_\alpha&
    \mathbf{a}_{\dot\beta}\mathbf{a}_\alpha{}\\[0.3em]
    \hdashline\\[-1.0em]
    -\bar{\mathbf{a}}_{\beta}\bar{\mathbf{a}}_{\dot\alpha}&
    \mathbf{a}_{\dot\beta}\bar{\mathbf{a}}_{\dot\alpha}\\
    \end{array}
  \right)
\end{align}
satisfying the $\mathfrak{gl}(n)$ algebra \eqref{eq:gln-alg}. The
element $\bar{\mathbf{C}}=\sum_{A=1}^{n}\bar{\mathbf{J}}_{AA}$ is
central. We denote the eigenspace of $\bar{\mathbf{C}}$ with
eigenvalue $c$ by $\bar{\mathcal{V}}_{c}\subset\mathcal{F}$. For each
$c\in\mathbb{Z}$ this space forms a unitary irreducible representation
of $\mathfrak{u}(p,q)$. The representation $\bar{\mathcal{V}}_c$ is
dual to $\mathcal{V}_{-c}$. It contains a highest weight state, which
is annihilated by all $\bar{\mathbf{J}}_{AB}$ with $A<B$. In case of
$q=0$ or $p=0$ the representation $\bar{\mathcal{V}}_{c}$ is a unitary
irreducible representation of $\mathfrak{u}(n)$. Having defined the
two classes of non-compact oscillator representations allows us to use
them at the sites of the monodromy $M(u)$ in \eqref{eq:mono}. At each
site we chose either a representation $\mathcal{V}_{c_i}$ with
generators $J_{AB}^i=\mathbf{J}_{AB}^i$ or $\bar{\mathcal{V}}_{c_i}$
with $J_{AB}^i=\bar{\mathbf{J}}_{AB}^i$. The monodromy $M(u)$, and
hence the representation of the Yangian, is completely specified by
$2N$ parameters, i.e.\ $N$ inhomogeneities $v_i\in\mathbb{C}$ and $N$
representation labels $c_i\in\mathbb{Z}$. We remark that the tensor
product decomposition of the oscillator representations employed at
the spin chain sites has been studied in \cite{Kashiwara:1978}, see
also e.g.\ \cite{Anderson:1968,Anderson:1968a} for exemplary results.

\section{Simple Sample Invariant}
\label{sec:simple}

Before formulating a Graßmannian integral for the just defined
oscillator representations of $\mathfrak{u}(p,q)$, it is instructive
to construct a simple solution of the Yangian invariance condition
\eqref{eq:def-yi} ``by hand''.

We consider a monodromy with two sites. To be able to construct a
$\mathfrak{gl}(n)$ singlet state, we choose for the first site a
``dual'' representation and for the second site an ``ordinary''
one. Hence the monodromy elements $M_{AB}(u)$ act on the space
$\bar{\mathcal{V}}_{c_1}\otimes\mathcal{V}_{c_2}$. The
$\mathfrak{gl}(n)$ generators, which appear in the Lax operators
\eqref{eq:lax} and consequently also in the Yangian generators
\eqref{eq:yangian-gen-explicit}, become
$J_{AB}^1=\bar{\mathbf{J}}_{AB}^1$ and
$J_{AB}^2=\mathbf{J}_{AB}^2$. To proceed we will make an ansatz for
the Yangian invariant state $|\Psi_{2,1}\rangle$, which is labeled by
the total number of sites $N=2$ and the number of ``dual'' sites
$K=1$. For this ansatz we introduce $\mathfrak{u}(p)$ and
$\mathfrak{u}(q)$ invariant contractions of oscillators, respectively,
\begin{align}
  \label{eq:two-site-inv}
  (1\bullet 2)
  =\sum_{\alpha=1}^p\bar{\mathbf{a}}^1_\alpha\bar{\mathbf{a}}^2_\alpha\,,\quad
  (1\circ 2)
  =\sum_{\dot\alpha=p+1}^{p+q}
  \bar{\mathbf{a}}^1_{\dot\alpha}\bar{\mathbf{a}}^2_{\dot\alpha}\,.
\end{align}
We assume $|\Psi_{2,1}\rangle$ to be a power series in $(1\bullet 2)$
and $(1\circ 2)$ acting on the Fock vacuum $|0\rangle$. Next, we
demand Yangian invariance \eqref{eq:def-yi} of this
ansatz. Furthermore, we impose that each site carries an irreducible
representation of $\mathfrak{u}(p,q)$, i.e.\
$\bar{\mathbf{C}}_1|\Psi_{2,1}\rangle=c_1|\Psi_{2,1}\rangle$ and
$\mathbf{C}_2|\Psi_{2,1}\rangle=c_2|\Psi_{2,1}\rangle$. This fixes the
invariant, up to a normalization constant, to be
\begin{align}
  \label{eq:nc21-sum}
  \begin{aligned}
  |\Psi_{2,1}\rangle&=2\pi i
  \sum_{\mathclap{\substack{g,h=0\\g-h=c_2+q}}}^\infty
  \frac{(1\bullet 2)^{g}}{g!}
  \frac{(1\circ 2)^{h}}{h!}
  |0\rangle
  =2\pi i\;
  \frac{I_{c_2+q}\big(2\sqrt{(1\bullet 2)(1\circ 2)}\big)}
  {\sqrt{(1\bullet 2)(1\circ 2)}^{\,c_2+q}}
  (1\bullet 2)^{c_2+q}|0\rangle\,,
  \end{aligned}
\end{align}
where we identified the sum with the series expansion of the modified
Bessel function of the first kind $I_\nu(x)$.\footnote{In the double
  sum in \eqref{eq:nc21-sum} $c_2+q$ can also manifestly take negative
  values. The validity of the Bessel function formulation in this case
  is easily verified using the series expansion.} The parameters of the
monodromy have to obey
\begin{align}
  \label{eq:nc21-para}
  v_1-v_2=1-n-c_2\,,\quad c_1=-c_2\in\mathbb{Z}\,.
\end{align}
We observe that the invariant \eqref{eq:nc21-sum} can be expressed as
a complex contour integral
\begin{align}
  \label{eq:nc21-int}
  |\Psi_{2,1}\rangle
  =\int\D C_{1 2}\frac{e^{C_{1 2}(1\bullet 2)+C^{-1}_{1 2}(1\circ 2)}|0\rangle}
  {C_{1 2}^{1+c_2+q}}\,.
\end{align}
Here the contour is a counterclockwise unit circle around the
essential singularity at $C_{12}=0$. It can be interpreted as group
manifold of the unitary group $U(1)$. The integral is easily evaluated
by using the residue theorem. This yields the series representation in
\eqref{eq:nc21-sum}. As we will see in the next section,
\eqref{eq:nc21-int} can already be considered as a simple Graßmannian
integral.

We finish this section with some remarks. The two-site invariant
\eqref{eq:nc21-sum} can be thought of as the oscillator analogue of
the twistor intertwiner that has been essential for the construction
of Yangian invariants in
\cite{Chicherin:2013ora,Kanning:2014maa,Broedel:2014pia}. This
intertwiner already appeared in the early days of twistor theory, cf.\
\cite{Penrose:1972ia,Hodges:1985ac}. We also note that recently a
two-site Yangian invariant for oscillator representations of
$\mathfrak{psu}(2,2|4)$ was used in \cite{Jiang:2014cya} based on a
construction in \cite{Alday:2005kq}. It takes the form of an
exponential function instead of a Bessel function as in
\eqref{eq:nc21-sum}.  This difference occurs because the invariant of
\cite{Jiang:2014cya} is not an eigenstate of the central element of
the symmetry algebra at each site.\footnote{We thank Ivan Kostov and
  Didina Serban for clarifying this point.} Furthermore, we remark
that employing the identity
\begin{align}
  \label{eq:hyper-bessel}
  \frac{I_\nu(2\sqrt{x})}{\sqrt{x}^{\,\nu}}
  =\frac{_0F_1(\nu+1;x)}{\Gamma(\nu+1)}\,,
\end{align}
cf.\ \cite{AbramowitzStegun:1964}, the invariant \eqref{eq:nc21-sum}
can alternatively be expressed in terms of a generalized
hypergeometric function $_0F_1(a,x)$. Sometimes this form is more
convenient because it avoids the ``spurious'' square roots, which are
absent in the series expansion. Additionally, the invariant in
\eqref{eq:nc21-sum} has infinite norm and thus is technically speaking
not an element of the Hilbert space
$\bar{\mathcal{V}}_{c_1}\otimes\mathcal{V}_{c_2}$. As a last aside,
let us consider the special case of the compact algebra
$\mathfrak{u}(p,0)$, i.e.\ we set $q=0$. The sum in
\eqref{eq:nc21-sum} simplifies to a single term
\begin{align}
  \label{eq:nc21-comp}
  |\Psi_{2,1}\rangle=2\pi i
  \frac{(1\bullet 2)^{c_2}}{c_2!}|0\rangle
\end{align}
with $c_2\geq 0$, where we used $(1\circ 2)^{h}=\delta_{0 h}$. This
form of the compact two-site Yangian invariant is known from
\cite{Frassek:2013xza}. 

\section{Graßmannian Integral Formula}
\label{sec:grassmannian}

At this point everything is set up to state our main formula, a
Graßmannian integral for Yangian invariants with oscillator
representations of the non-compact algebra $\mathfrak{u}(p,q)$. We
motivate it by combining our knowledge of the Graßmannian integral for
scattering amplitudes \eqref{eq:grass_amp} with that of the simple
sample invariant \eqref{eq:nc21-int}.  In this section we merely state
the resulting formula. A proof of its Yangian invariance is deferred
to appendix~\ref{sec:proof}.

A Yangian invariant for a monodromy with $N=2K$ sites, out of which
the first $K$ are ``dual'' sites and the remaining $K=N-K$ sites are
``ordinary'', is given by the \emph{Graßmannian integral formula}
\begin{align}
  \label{eq:grass-int}
  |\Psi_{N,K}\rangle
  =\int\D\mathcal{C}
  \frac{e^{\tr(\mathcal{C}\mathbf{I}_\bullet^t+\mathbf{I}_\circ \mathcal{C}^{-1})}
    |0\rangle}
  {(\det\mathcal{C})^q(1,\ldots,K)^{1+v_{K}^+-v_1^-}
    \cdots (N,\ldots,K-1)^{1+v_{K-1}^+-v_N^-}}\,.
\end{align}
Here the numerator can be understood as a matrix generalization of the
sample invariant \eqref{eq:nc21-int}. The single contractions of
oscillators in the exponent are replaced by the matrices
\begin{align}
  \label{eq:osc-matrix}
  \mathbf{I}_{\mathrel{\ooalign{\raisebox{0.4ex}{$\scriptstyle\bullet$}\cr\raisebox{-0.4ex}{$\scriptstyle\circ$}}}}=
  \begin{pmatrix}
    (1\mathrel{\ooalign{\raisebox{0.7ex}{$\bullet$}\cr\raisebox{-0.3ex}{$\circ$}}} K+1)&
    \cdots&
    (1\mathrel{\ooalign{\raisebox{0.7ex}{$\bullet$}\cr\raisebox{-0.3ex}{$\circ$}}} N)\\
    \vdots&&\vdots\\
    (K\mathrel{\ooalign{\raisebox{0.7ex}{$\bullet$}\cr\raisebox{-0.3ex}{$\circ$}}} K+1)&
    \cdots&
    (K\mathrel{\ooalign{\raisebox{0.7ex}{$\bullet$}\cr\raisebox{-0.3ex}{$\circ$}}} N)\\
  \end{pmatrix}.
\end{align}
These $K\times K$ matrices $\mathbf{I}_\bullet$ and $\mathbf{I}_\circ$
contain, respectively, all possible $\mathfrak{u}(p)$ and
$\mathfrak{u}(q)$ invariant contractions of the type
\eqref{eq:two-site-inv} between a ``dual'' and an ``ordinary''
site. The denominator of \eqref{eq:grass-int} is analogous to the
Graßmannian integral for scattering amplitudes \eqref{eq:grass_amp}
and contains the minors of the $K\times N$ matrix $C$ defined in
\eqref{eq:grassm-matrix}. Notice however the extra factor of
$(\det\mathcal{C})^q$. The gauge fixing of the matrix $C$ corresponds
to the order of ``dual'' and ``ordinary'' sites. Furthermore, the
measure is the same as in \eqref{eq:grass_amp}. Finally, the $2N$
parameters $\{v_i^+,v_i^-\}$ have to obey the $N$ relations in
\eqref{eq:spec-perm}.

Next, we specify in detail the monodromy $M(u)$ with which the
Graßmannian integral for $|\Psi_{N,K}\rangle$ in \eqref{eq:grass-int}
satisfies the Yangian invariance condition \eqref{eq:def-yi}. The
elements $M_{AB}(u)$ of this monodromy act on the space
$\bar{\mathcal{V}}_{c_1}\otimes\cdots\otimes\bar{\mathcal{V}}_{c_K}
\otimes\mathcal{V}_{c_{K+1}}\otimes\cdots\otimes\mathcal{V}_{c_N}$.
The $\mathfrak{gl}(n)$ generators in the Lax operators \eqref{eq:lax}
and in the Yangian generators \eqref{eq:yangian-gen-explicit} become
\begin{align}
  \label{eq:gen-mono}
  J^i_{AB}=
  \begin{cases}
    \bar{\mathbf{J}}_{AB}^i\quad\text{for}\quad i=1,\ldots,K\,,\\
    \mathbf{J}_{AB}^i\quad\text{for}\quad i=K+1,\ldots,N\,.
  \end{cases}
\end{align}
In the formula \eqref{eq:grass-int}, the $2N$ parameters $\{v_i,c_i\}$
describing the monodromy were traded for a different set of $2N$
parameters $\{v_i^+,v_i^-\}$. They are related by, cf.\
\cite{Beisert:2014qba},\footnote{This redefinition of parameters has
  also been discussed in \cite{Kanning:2014maa} for the
  $\mathfrak{u}(2)$ case, i.e.\ $n=2$. The equation for $v_i'$ differs
  from the corresponding equation (40) in \cite{Kanning:2014maa} by a
  shift of $1$ at the dual sites. This shift originates from a shift
  of the inhomogeneities of the Lax operators at those sites.}
\begin{align}
  \label{eq:spec-redef}
  v^\pm_i=v_i'\pm\frac{c_i}{2}\,,\quad
  v_i'=v_i-\frac{c_i}{2}+
  \begin{cases}
    n-1&\text{for}\quad i=1,\ldots K\,,\\
    0&\text{for}\quad i=K+1,\ldots N\,.
  \end{cases}
\end{align}
The monodromy is equivalently described by either $\{v_i,c_i\}$ or
$\{v_i^+,v_i^-\}$. Notice, however, that for the oscillator
representations under consideration the deformation parameters
$v_i^\pm$ \emph{cannot} be any complex numbers. They have to be such
that the corresponding $c_i$ are integers. This completes the
specification of the monodromy.

Let us remark that imposing the condition $N=2K$ guarantees
$\mathcal{C}$ to be a square matrix. Thus it is sensible to use its
inverse in \eqref{eq:grass-int}. In the compact special case
$\mathfrak{u}(p,0)$ we have $\mathbf{I}_\circ=0$, thus
$\mathcal{C}^{-1}$ is absent from \eqref{eq:grass-int} and the
Graßmannian integral yields Yangian invariants also for $N\neq
2K$. However, we do not elaborate on the compact case in this work.
We note that because of $\mathbf{I}_\circ=0$, the \emph{compact} case
of \eqref{eq:grass-int} is reminiscent of the link representation of
scattering amplitudes, cf.\ \cite{ArkaniHamed:2009dn}. It is different
though, as the amplitudes transform under the \emph{non-compact}
algebra $\mathfrak{psu}(2,2|4)$. Another remark concerns the
multi-dimensional contour of integration in \eqref{eq:grass-int},
which we did not specify so far. The proof in appendix \ref{sec:proof}
only assumes that the boundary terms vanish upon integration by parts,
which is satisfied in particular for closed contours. The choice of
the integration contour will be an issue in the following sections.

\section{Unitary Matrix Models}
\label{sec:matrix-models}

In this section we choose a ``unitary contour'' and special values of
the deformation parameters $v_i^\pm$ in the Graßmannian integral
\eqref{eq:grass-int}. Thereby this integral reduces to the
Brezin-Gross-Witten matrix model or even a slight generalization
thereof, the Leutwyler-Smilga model. In this special case, the
Graßmannian integral can be computed easily by applying well
established matrix model techniques. In this way, we obtain a
representation of these Yangian invariants in terms of Bessel
functions.

In order to reduce \eqref{eq:grass-int} with $N=2K$ to the
Leutwyler-Smilga integral, we restrict to a special solution of the
constraints in \eqref{eq:spec-perm} on the deformation parameters
$v_i^\pm$. The solution has to be such that all minors in
\eqref{eq:grass-int}, except for $(1,\ldots,K)=1$ and
$(N-K+1,\ldots,N)=\det\mathcal{C}$, have a vanishing exponent. A short
calculation shows that this solution depends only on two parameters
$v\in\mathbb{C}, c\in\mathbb{Z}$. It is given by
\begin{align}
  \label{eq:ls-int-para}
  \begin{aligned}
    v_i&=v-c-n+1+(i-1)\,,\quad&c_i&=-c &\quad&\text{for}\quad i=1,\ldots,K\,,\\
    v_i&=v+(i-K-1)\,,\quad&c_i&=\phantom{-}c &\quad&\text{for}\quad i=K+1,\ldots,2K\,.\\
  \end{aligned}
\end{align}
Here we used \eqref{eq:spec-redef} to change from the variables
$\{v_i^+,v_i^-\}$ employed in \eqref{eq:grass-int} to the variables
$\{v_i,c_i\}$. Let us now focus on the measure
$\D\mathcal{C}=\bigwedge_{k,l}\D C_{k,l}$ in \eqref{eq:grass-int}. One
readily verifies that
\begin{align}
  \label{eq:measure-haar}
  [\D\mathcal{C}]=\chi\,\frac{\D\mathcal{C}}{(\det\mathcal{C})^K}\,,
\end{align}
with a constant number $\chi\in\mathbb{C}$, is invariant under
$\mathcal{C}\mapsto\mathcal{V}\mathcal{C}$ and
$\mathcal{C}\mapsto\mathcal{C}\mathcal{V}$ for any constant matrix
$\mathcal{V}\in GL(K)$. Hence for unitary $\mathcal{C}$ the expression
$[\D\mathcal{C}]$ defined in \eqref{eq:measure-haar} is the Haar
measure on the unitary group $U(K)$. The normalization $\chi$ is
chosen such that $\int_{U(K)}[\D\mathcal{C}]= 1$. We select a
``unitary contour'' in the Graßmannian integral \eqref{eq:grass-int}
by demanding $\mathcal{C}^\dagger=\mathcal{C}^{-1}$. This allows us to
express the Yangian invariant with the special choice of deformation
parameters \eqref{eq:ls-int-para} as
\begin{align}
  \label{eq:red-matrix-fin-inv}
  |\Psi_{2K,K}\rangle
  =\chi^{-1}\int_{U(K)}[\D\mathcal{C}]
  \frac{e^{\tr(\mathcal{C}\mathbf{I}_\bullet^t+\mathbf{I}_\circ \mathcal{C}^\dagger)}|0\rangle}
  {(\det\mathcal{C})^{c+q}}\,,
\end{align}
where $c\in\mathbb{Z}$ is a free
parameter. Eq.~\eqref{eq:red-matrix-fin-inv} is known as
\emph{Leutwyler-Smilga model} \cite{Leutwyler:1992yt}, where the
matrices $\mathbf{I}_\bullet^t$ and $\mathbf{I}_\circ$ are considered
as sources. For $c=-q$ it becomes the \emph{Brezin-Gross-Witten model}
\cite{Gross:1980he,Brezin:1980rk}. Remarkably, the integral
\eqref{eq:red-matrix-fin-inv} can be computed exactly. For two
\emph{independent} source matrices $\mathbf{I}_\bullet^t$ and
$\mathbf{I}_\circ$ this was achieved in \cite{Schlittgen:2002tj} using
the character expansion methods of \cite{Balantekin:2000vn},
\begin{align}
  \label{eq:ls-integral-bessel}
  |\Psi_{2K,K}\rangle=
  \chi^{-1}\prod_{j=0}^{K-1}j!
  \frac{(\det\mathbf{I}_\bullet^t)^{c+q}}
  {\Delta(\mathbf{I}_\circ\mathbf{I}_\bullet^t)}
  \det\left(\frac{I_{k+c+q-K}\big(2\sqrt{(\mathbf{I}_\circ\mathbf{I}_\bullet^t)_l}\big)}
    {\sqrt{(\mathbf{I}_\circ\mathbf{I}_\bullet^t)_l}^{\,k+c+q-K}}\right)_{k,l}
  |0\rangle\,.
\end{align}
Assuming the matrix $\mathbf{I}_\circ\mathbf{I}_\bullet^t$ to be
diagonalizable, we denote its $l$-th eigenvalue by
$(\mathbf{I}_\circ\mathbf{I}_\bullet^t)_l$. Furthermore,
$\Delta(\mathbf{I}_\circ\mathbf{I}_\bullet^t)=\det
({(\mathbf{I}_\circ\mathbf{I}_\bullet^t)_l}^{k-1})_{k,l}$ is the
Vandermonde determinant. The formula \eqref{eq:ls-integral-bessel}
involving a determinant of Bessel functions generalizes the single
Bessel function that we found for the sample Yangian invariant
$|\Psi_{2,1}\rangle$ in \eqref{eq:nc21-sum}.

In this section we showed that the choice of a ``unitary contour'' in
the Graßmannian integral \eqref{eq:grass-int} is appropriate for the
\emph{special} deformation parameters $v_i^\pm$ given by
\eqref{eq:ls-int-para}. We conjecture that this contour can also be
used for the Graßmannian integral \eqref{eq:grass-int} with
\emph{general} deformation parameters. In this case one is lead to a
novel unitary matrix model of the type \eqref{eq:red-matrix-fin-inv}
containing powers of principal minors of the matrix $\mathcal{C}$ in
addition to $\det{\mathcal{C}}$. In the next section we illustrate for
a non-trivial example that this model indeed produces the correct
Yangian invariant.

\section{Another Sample Invariant: R-Matrix}
\label{sec:r-matrix}

Let us now apply the ``unitary contour'' to the Graßmannian integral
\eqref{eq:grass-int} for the sample invariant
$|\Psi_{4,2}\rangle$. This invariant is of special importance because
its Yangian invariance condition \eqref{eq:def-gen} can be translated
into the Yang-Baxter equation, cf.\ \cite{Frassek:2013xza}. Therefore
$|\Psi_{4,2}\rangle$ is equivalent to the $\mathfrak{u}(p,q)$
R-matrix.

We begin by choosing $\mathcal{C}$ to be unitary which transforms the
integral \eqref{eq:grass-int} into
\begin{align}
  \label{eq:nc42-int}
  |\Psi_{4,2}\rangle=\chi^{-1}
  \int_{U(2)}[\D\mathcal{C}]
  \frac{\,
    e^{\tr(\mathcal{C}\mathbf{I}_\bullet^t+\mathbf{I}_\circ \mathcal{C}^{\dagger})}|0\rangle}
  {(-C_{1 3})^{1+z}(\det{\mathcal{C}})^{-1+q-z+c_3}(-C_{2 4})^{1+z-c_3+c_4}}\,
\end{align}
with the abbreviation $z=v_3-v_4$. The constraints on the deformation
parameters in \eqref{eq:spec-perm} read explicitly
\begin{gather}
  \label{eq:nc42-perm}
    v_1-v_3=1-n-c_3\,,\quad
    c_1=-c_3\in\mathbb{Z}\,,\quad
    v_2-v_4=1-n-c_4\,,\quad
    c_2=-c_4\in\mathbb{Z}\,.
\end{gather}
Notice that \eqref{eq:nc42-int} is a generalization of the
Leutwyler-Smilga model \eqref{eq:red-matrix-fin-inv}, as it contains
in addition the principal minors $C_{13}$ and $C_{24}$ of the unitary
$2\times 2$ matrix $\mathcal{C}$.  This currently hinders the direct
application of matrix model techniques to evaluate
\eqref{eq:nc42-int}. Therefore we resort to an explicit
parameterization,
\begin{align}
  \label{eq:nc42-para}
  \mathcal{C}=
  \begin{pmatrix}
    C_{13}&C_{14}\\
    C_{23}&C_{24}\\
  \end{pmatrix}
  =c
  \begin{pmatrix}
    a \cos\theta &-b \sin\theta \\
    b^{-1} \sin\theta &a^{-1} \cos\theta \\
  \end{pmatrix},
\end{align}
where $\theta\in[0,\frac{\pi}{2}]$ and $a=e^{i\alpha}, b=e^{i\beta},
c=e^{i\gamma}$ with $\alpha,\beta\in[0,2\pi]$ and
$\gamma\in[0,\pi]$. With this the Haar measure \eqref{eq:measure-haar}
becomes
\begin{align}
  \label{eq:haar-u2}
  [\D\mathcal{C}]=\chi\frac{4\sin{\theta}\cos{\theta}}{abc}
  \D a\wedge\D b\wedge\D c\wedge\D\theta\,.
\end{align}
We observe that the exponents of $a,b,c$ in denominator of
\eqref{eq:nc42-int} combine into integers, where for the moment we
ignore that this rearrangement is not allowed for generic values of
the exponent $z\in\mathbb{C}$. Thus the integrals in the variables
$a,b,c$ can be performed by means of the residue theorem,
\begin{align}
  \label{eq:nc42-sum-euler}
  \begin{aligned}
    |\Psi_{4,2}\rangle=
    (-1)^{c_4-c_3}(2\pi i)^3\;
    \,\smash{\sum_{\mathclap{\substack{g_{13},\ldots,g_{24}=0\\h_{13},\ldots,h_{24}=0\\\text{with \eqref{eq:nc42-sum-constraints}}}}}^\infty}\,\quad\quad
    &\frac{(1\bullet 3)^{g_{13}}}{g_{13}!}\frac{(1\bullet 4)^{g_{14}}}{g_{14}!}
    \frac{(2\bullet 3)^{g_{23}}}{g_{23}!}\frac{(2\bullet 4)^{g_{24}}}{g_{24}!}\\
    \cdot\,&\frac{(1\circ 3)^{h_{13}}}{h_{13}!}\frac{(1\circ 4)^{h_{14}}}{h_{14}!}
    \frac{(2\circ 3)^{h_{23}}}{h_{23}!}\frac{(2\circ 4)^{h_{24}}}{h_{24}!}
    |0\rangle\\
    \cdot\,(-1)^{g_{14}+h_{14}}&\text{B}(g_{14}+h_{23}+1,h_{13}+g_{24}-z+c_3-c_4)\,.
  \end{aligned}
\end{align}
In this formula the constraints
\begin{align}
  \label{eq:nc42-sum-constraints}
  \begin{aligned}
    g_{13}-h_{13}+g_{14}-h_{14}&=-c_1+q\,,&\quad
    g_{23}-h_{23}+g_{24}-h_{24}&=-c_2+q\,,\\
    g_{13}-h_{13}+g_{23}-h_{23}&=\phantom{-}c_3+q\,,&\quad
    g_{14}-h_{14}+g_{24}-h_{24}&=\phantom{-}c_4+q\,\\
  \end{aligned}
\end{align}
on the summation range guarantee that $|\Psi_{4,2}\rangle$ is an
eigenstate of $\bar{\mathbf{C}}^1, \bar{\mathbf{C}}^2, \mathbf{C}^3,
\mathbf{C}^4$ with eigenvalues $c_1, c_2, c_3, c_4$, respectively. The
remaining $\theta$-integration yields the Euler beta function
\begin{align}
  \label{eq:beta}
  \text{B}(x,y)=2\int_{0}^{\frac{\pi}{2}}
  \D \theta (\sin\theta)^{2x-1}(\cos\theta)^{2y-1}
  =\frac{\Gamma(x)\Gamma(y)}{\Gamma(x+y)}\,,
\end{align}
which is valid for $\text{Re}\,x, \text{Re}\,y>0$, cf.\
\cite{AbramowitzStegun:1964}. This means $c_3-c_4>\text{Re}\,z$ for
the arguments of the beta function in \eqref{eq:nc42-sum-euler}. The
expression \eqref{eq:nc42-sum-euler} is our final form of the Yangian
invariant $|\Psi_{4,2}\rangle$, i.e.\ the $\mathfrak{u}(p,q)$ R-matrix
for oscillator representations. The parameter $z$ is the spectral
parameter of this R-matrix. A formula analogous to
\eqref{eq:nc42-sum-euler}, however derived in a completely different
way, can be obtained by specializing the $\mathfrak{u}(p,q|r)$
R-matrix expression found in \cite{Ferro:2013dga} to the bosonic
case. At this point we remark that the integrand in
\eqref{eq:nc42-int} is multi-valued for generic $z$ and thus the
$U(2)$ contour is not closed. Hence in principle the formal proof in
appendix~\ref{sec:proof} does not directly apply. Therefore we
verified the Yangian invariance of $|\Psi_{4,2}\rangle$ explicitly on
the level of the series expansion \eqref{eq:nc42-sum-euler}. Finally,
it is worth noting that in the compact case
$\mathfrak{u}(p,0)=\mathfrak{u}(n)$ the invariant
\eqref{eq:nc42-sum-euler} simplifies to
\begin{align}
  \label{eq:nc42-compact}
  \begin{aligned}
    |\Psi_{4,2}\rangle=
    (-1)^{c_4-c_3}2(2\pi i)^3
    \sum_{g_{14}=0}^\infty
    &\frac{(1\bullet 3)^{c_3-g_{14}}}{(c_3-g_{14})!}\frac{(1\bullet 4)^{g_{14}}}{g_{14}!}
    \frac{(2\bullet 3)^{g_{14}}}{g_{14}!}\frac{(2\bullet 4)^{c_4-g_{14}}}{(c_4-g_{14})!}|0\rangle\\
    \cdot\,&(-1)^{g_{14}}\text{B}(g_{14}+1,-z+c_3-g_{14})\,.
  \end{aligned}
\end{align}
This agrees with the compact invariant $|\Psi_{4,2}\rangle$ obtained
in \cite{Frassek:2013xza} up to a normalization factor.

\section{Conclusions and Outlook}
\label{sec:conclusions}

In this work we showed that the Graßmannian integral, commonly used in
the realm of $\mathcal{N}=4$ SYM scattering amplitudes, can be applied
to construct Yangian invariants for oscillator representations of the
non-compact algebra $\mathfrak{u}(p,q)$. We found that in this setting
the integral takes the form of a matrix model which generalizes the
Brezin-Gross-Witten and the Leutwyler-Smilga model. Our results also
imply that these two well-known matrix models are Yangian invariant in
the external source fields!

Our work calls for a series of further investigations, both on a
technical and on a conceptual level. Technically, the generalization
to superalgebras $\mathfrak{u}(p,q|r)$ should impose no
obstacles. This is of importance to cover the $\mathfrak{psu}(2,2|4)$
case relevant for amplitudes. In addition, the applicability of the
``unitary contour'' has to be investigated further. In particular,
replacing $\mathcal{C}^{-1}$ by the conjugate transpose
$\mathcal{C}^\dagger$ in the Graßmannian integral formula
\eqref{eq:grass-int} should also provide a way to avoid the ``split
helicity'' constraint $N=2K$. Here the issue is to use an appropriate
measure on the complex Stiefel manifold of \emph{rectangular}
$K\times(N-K)$ matrices $\mathcal{C}$ with
$\mathcal{C}\mathcal{C}^\dagger=1_{K\times K}$, see e.g.\
\cite{Mathai:1997}. This generalizes the unitary group manifold to the
case of rectangular matrices. Moreover, we want to apply matrix model
technology for the evaluation of the Graßmannian integral
\eqref{eq:grass-int} beyond the case of the Leutwyler-Smilga model
\eqref{eq:red-matrix-fin-inv}. One might wonder whether the Bessel
function formula \eqref{eq:ls-integral-bessel} generalizes to the case
of Yangian invariants with general deformation parameters
$v_i^\pm$. This formula would include the R-matrix constructed ``by
hand'' in section~\ref{sec:r-matrix}. One promising technique for this
endeavor is a character expansion, which was successfully employed for
the Leutwyler-Smilga model \eqref{eq:red-matrix-fin-inv}, see
\cite{Schlittgen:2002tj,Balantekin:2000vn}. Another auspicious method
is the use of Gelfand-Tzetlin coordinates, which has been applied to
compute correlation functions of the Itzykson-Zuber model
\cite{Shatashvili:1992fw}. In our setting these coordinates might be
well adapted to the minors appearing in the Graßmannian integral
\eqref{eq:grass-int}. A further interesting point to be addressed in
the future is the precise relation between the Graßmannian integral
for twistors and that for oscillator representations
\eqref{eq:replace}. There should exist a change of basis transforming
the delta function of twistors into the exponential function of
oscillators. A twistorial description of the $\mathfrak{u}(p,q)$
oscillator representations, a.k.a.\ ``ladder representations'', is
discussed e.g.\ in \cite{Dunne:1990}.

Even more exciting questions arise on the conceptual level. It is well
known that matrix models possess an integrable structure, see e.g.\
\cite{Morozov:1994hh} and references therein. Their partition
functions, like e.g.\ \eqref{eq:red-matrix-fin-inv}, correspond to
solutions, so-called $\tau$-functions, of classically integrable
hierarchies. There should be a relation between this \emph{classical}
integrable structure and \emph{quantum} integrability in the sense of
Yangian invariance. One might even ask if there is an integrable
hierarchy governing (tree-level) $\mathcal{N}=4$ SYM scattering
amplitudes. Finally, let us speculate that our matrix model approach
might also provide a conceptually clear route to loop-amplitudes. The
$\mathfrak{psu}(2,2|4)$ analogues of the oscillator representations,
which we are using in this work, feature prominently in the spectral
problem of $\mathcal{N}=4$ SYM. There it is understood how to
introduce the coupling constant of the theory as a central extension
of the algebra. Appealing to a common integrable structure of the
$\mathcal{N}=4$ model, we suspect that in the oscillator basis such a
coupling can also be introduced in the Graßmannian integral.

\section*{Acknowledgments}
\label{sec:acknowledgements}
We thank Shota Komatsu, Ivan Kostov, Carlo Meneghelli and Didina
Serban for helpful discussions. We are thankful to the CERN Theory
Division for their hospitality during the initial phase of this
project. We also gratefully acknowledge support from the Simons Center
for Geometry and Physics, Stony Brook University, and of the
C.~N.~Yang Institute for Theoretical Physics, where some of the
research for this paper was performed.  N.~K. and Y.~K. acknowledge
the support of the Marie Curie International Research Staff Exchange
Network UNIFY of the European Union’s Seventh Framework Programme
FP7-People-2010-IRSES under grant agreement No.\ 269217, which allowed
them to visit Stony Brook University. This research is supported in
part by the SFB 647 \emph{``Raum-Zeit-Materie. Analytische und
  Geometrische Strukturen''} and the Marie Curie Network GATIS
(\texttt{\href{http://gatis.desy.eu}{gatis.desy.eu}}) of the European
Union’s Seventh Framework Programme FP7-2007-2013 under grant
agreement No.\ 317089. N.~K. is supported by a Ph.D. scholarship of
the International Max Planck Research School for Geometric Analysis,
Gravitation and String Theory.

\appendix

\section{Proof of Yangian Invariance}
\label{sec:proof}

In this appendix we prove the Yangian invariance \eqref{eq:def-yi-gen}
of the Graßmannian integral \eqref{eq:grass-int} for the invariant
$|\Psi_{N, K}\rangle$ with $N=2K$ sites and representations of the
non-compact algebra $\mathfrak{u}(p,q)$. With straightforward
modifications this proof also applies to the compact case, i.e.\
$q=0$, where, in particular, $\mathbf{I}_\circ=0$ and $N\neq 2K$ is
possible.

Let us start with the ansatz
\begin{align}
  \label{eq:osc-d-matrix}
  e^{\tr(\mathcal{C}\mathbf{I}_\bullet^t+\mathbf{I}_\circ \mathcal{C}^{-1})}|0\rangle\,,
\end{align}
which we recognize as the exponential function in
\eqref{eq:grass-int}. We want to show that this ansatz satisfies the
first equation of \eqref{eq:def-yi-gen}, that is to say
$\mathfrak{gl}(n)$ invariance. With the $\mathfrak{gl}(n)$ generators
of our monodromy defined in \eqref{eq:gen-mono}, the Yangian
generators appearing in this equation read
\begin{align}
  \label{eq:level1-dual-normal}
  M^{(1)}_{AB}=\sum_{k=1}^K \bar{\mathbf{J}}^k_{BA}+\sum_{l=K+1}^N \mathbf{J}^l_{BA}\,.
\end{align}
To evaluate the action of this operator on the ansatz
\eqref{eq:osc-d-matrix} we compute
\begin{align}
  \label{eq:action-level-one}
  \begin{aligned}
    (\bar{\mathbf{J}}^k_{AB})\,e^{\tr(\mathcal{C}\mathbf{I}_\bullet^t+\mathbf{I}_\circ \mathcal{C}^{-1})}|0\rangle
    &=
    \left(
      \begin{array}{c:c}
        -\sum_w\bar{\mathbf{a}}^w_\alpha \,\bar{\mathbf{a}}^k_\beta \,C_{k w}&
        \sum_{w,w'}\bar{\mathbf{a}}^w_\alpha\,\bar{\mathbf{a}}^{w'}_{\dot\beta} D_{w' k}C_{k w}\\[0.3em]
        \hdashline\\[-1.0em]
        -\bar{\mathbf{a}}^k_{\dot\alpha}\,\bar{\mathbf{a}}^k_\beta&
        \sum_w\bar{\mathbf{a}}^k_{\dot\alpha}\,\bar{\mathbf{a}}^w_{\dot\beta}\,D_{w k}+\delta_{\dot\alpha\dot\beta}\\
      \end{array}
    \right)
    e^{\tr(\mathcal{C}\mathbf{I}_\bullet^t+\mathbf{I}_\circ \mathcal{C}^{-1})}|0\rangle\,,\\ 
    (\mathbf{J}^l_{AB})\,e^{\tr(\mathcal{C}\mathbf{I}_\bullet^t+\mathbf{I}_\circ \mathcal{C}^{-1})}|0\rangle
    &=
    \left(
      \begin{array}{c:c}
        \sum_v\bar{\mathbf{a}}^l_\alpha\,\bar{\mathbf{a}}^v_\beta \,C_{v l}&
        -\bar{\mathbf{a}}^l_{\alpha}\,\bar{\mathbf{a}}^l_{\dot\beta}\\[0.3em]
        \hdashline\\[-1.0em]
        \sum_{v,v'}\bar{\mathbf{a}}^v_{\dot\alpha}\,\bar{\mathbf{a}}^{v'}_\beta C_{v' l}\,D_{l v}&
        -\sum_v\bar{\mathbf{a}}^v_{\dot\alpha}\,\bar{\mathbf{a}}^l_{\dot\beta}\,D_{l v}-\delta_{\dot\alpha\dot\beta}\\
      \end{array}
    \right)
    e^{\tr(\mathcal{C}\mathbf{I}_\bullet^t+\mathbf{I}_\circ \mathcal{C}^{-1})}|0\rangle\,,
  \end{aligned}
\end{align}
where the components of the matrix $\mathcal{C}^{-1}$ are denoted by
$D_{lk}$. Here and in the rest of this proof the indices $k,v,v'$
always take the values $1,\ldots,K$ while $l,w,w'$ are in the range
$K+1,\ldots,N$. Now one immediately obtains
\begin{align}
  \label{eq:level-one}
  M^{(1)}_{AB}\,
  e^{\tr(\mathcal{C}\mathbf{I}_\bullet^t+\mathbf{I}_\circ \mathcal{C}^{-1})}\,|0\rangle=0\,.
\end{align}
Hence the first equation of \eqref{eq:def-yi-gen} holds for the ansatz
\eqref{eq:osc-d-matrix}.

However, each site of the ansatz \eqref{eq:osc-d-matrix} does not yet
transform in an irreducible representation of the algebra
$\mathfrak{u}(p,q)$. In fact, \eqref{eq:osc-d-matrix} is not an
eigenstate of the central elements
$\mathbf{C}^l=\sum_{A=1}^n\mathbf{J}_{AA}^l$ and
$\bar{\mathbf{C}}^k=\sum_{A=1}^n\bar{\mathbf{J}}_{AA}^k$ that were
defined in the context of \eqref{eq:osc-sym-gen} and
\eqref{eq:osc-dual-gen}, respectively. To obtain eigenstates we have
to pick special linear combinations of the ansatz
\eqref{eq:osc-d-matrix},
\begin{align}
  \label{eq:ansatz-general}
  |\Psi_{N, K}\rangle=\int\D\mathcal{C}\,f(\mathcal{C})\,
  e^{\tr(\mathcal{C}\mathbf{I}_\bullet^t+\mathbf{I}_\circ \mathcal{C}^{-1})}|0\rangle\,.
\end{align}
It turns out to be suitable to choose an integrand that contains only
consecutive minors of the matrix $C$ defined in
\eqref{eq:grassm-matrix},
\begin{align}
  \label{eq:ansatz-minors}
  f(\mathcal{C})=\frac{1}{(1,\ldots,K)^{1+\alpha_1}\cdots(N,\ldots,K-1)^{1+\alpha_N}}\,
\end{align}
with arbitrary complex constants $\alpha_i$. With this integrand the
ansatz \eqref{eq:ansatz-general} is an eigenstate of the central
elements,
\begin{align}
  \label{eq:ccbar-eval}
  \bar{\mathbf{C}}^k\,|\Psi_{N, K}\rangle
  =\biggl(\,\,q-\sum_{\mathclap{i=k+1}}^{\mathclap{k+N-K}}\alpha_i\,\,\biggr)\,|\Psi_{N, K}\rangle\,,\quad
  \mathbf{C}^l\,|\Psi_{N, K}\rangle
  =\biggl(-q+\sum_{\mathclap{i=l-K+1}}^l\alpha_i\,\,\biggr)\,|\Psi_{N, K}\rangle\,.
\end{align}
To show this property we assumed that upon integration by parts the
boundary terms vanish. Furthermore, we employed the identity
\begin{align}
  \label{eq:diff-ansatz}
  \frac{\D}{\D C_{k l}}e^{\tr(\mathcal{C}\mathbf{I}_\bullet^t+\mathbf{I}_\circ \mathcal{C}^{-1})}|0\rangle=
  \biggl((k\bullet l)-\sum_{v,w}D_{w k}D_{l v}(v\circ w)\biggr)
  e^{\tr(\mathcal{C}\mathbf{I}_\bullet^t+\mathbf{I}_\circ \mathcal{C}^{-1})}|0\rangle\,,
\end{align}
which is easily verified taking into account $\frac{\D}{\text{d} C_{k
    l}}D_{w v}=-D_{w k}D_{l v}$. In addition, in evaluating
derivatives of the minors in $f(\mathcal{C})$ we used, cf.\
\cite{Drummond:2010qh,Drummond:2010uq},
\begin{align}
  \label{eq:deriv-minor}
  \sum_{w} C_{k w} \frac{\D}{\D C_{k w}} (i, \ldots, i+K-1)^{1+\alpha_i} 
  = (1+\alpha_i )\,(i, \ldots, i+K-1)^{1+\alpha_i}\,
\end{align}
for $i = k+1, \ldots, k+N-K$. For other values of $i$ the left hand
side in \eqref{eq:deriv-minor} vanishes due to the gauge fixing of $C$
in \eqref{eq:grassm-matrix}.

Next, we turn our attention to the second equation of the Yangian
invariance condition \eqref{eq:def-yi-gen}, which involves the
generators $M_{AB}^{(2)}$. From \eqref{eq:def-gen} with $r=2$ and
$s=1$ one sees that if a state $|\Psi\rangle$ is annihilated by all
$M_{AB}^{(1)}$ and by one of the generators $M_{AB}^{(2)}$, e.g.\ by
$M_{11}^{(2)}$, then it is annihilated by all $M_{AB}^{(2)}$. Thus in
our case it is sufficient to verify the second equation of
\eqref{eq:def-yi-gen} for one of the four blocks of generators, say
for $M^{(2)}_{\alpha\beta}$. Expressions for these generators can be
found in \eqref{eq:yangian-gen-explicit}. We compute the action of all
terms appearing therein on our ansatz \eqref{eq:osc-d-matrix},
\begin{align}
  \label{eq:jj-action}
  \begin{aligned}
    &\sum_{I=1}^n \,\bar{\mathbf{J}}^k_{I\alpha}\mathbf{J}_{\beta I}^l\,
    e^{\tr(\mathcal{C}\mathbf{I}_\bullet^t+\mathbf{I}_\circ \mathcal{C}^{-1})}|0\rangle
    =-\bar{\mathbf{a}}^k_\alpha\,\bar{\mathbf{a}}^l_\beta
    \biggl(\,\sum_{v,w}\,C_{v l}C_{k w}\,\frac{\D}{\D C_{v w}}+p\,C_{k l}\biggr)
    e^{\tr(\mathcal{C}\mathbf{I}_\bullet^t+\mathbf{I}_\circ \mathcal{C}^{-1})}|0\rangle\,,\\
    &\sum_{I=1}^n\,\bar{\mathbf{J}}^k_{I\alpha}\bar{\mathbf{J}}_{\beta I}^{k'}\,
    e^{\tr(\mathcal{C}\mathbf{I}_\bullet^t+\mathbf{I}_\circ \mathcal{C}^{-1})}|0\rangle
    =
    \sum_{w,w'}\bar{\mathbf{a}}^k_\alpha\,\bar{\mathbf{a}}^w_\beta\, C_{k' w}\,C_{k w'}\frac{\D}{\D C_{k' w'}}
    e^{\tr(\mathcal{C}\mathbf{I}_\bullet^t+\mathbf{I}_\circ \mathcal{C}^{-1})}|0\rangle\,,\\
    &\sum_{I=1}^n\,\mathbf{J}^l_{I\alpha}\mathbf{J}_{\beta I}^{l'}\,
    e^{\tr(\mathcal{C}\mathbf{I}_\bullet^t+\mathbf{I}_\circ \mathcal{C}^{-1})}|0\rangle
    =
    \sum_{v,v'}\bar{\mathbf{a}}^v_\alpha\,\bar{\mathbf{a}}^{l'}_\beta\, C_{v l}\,C_{v' l'}\,\frac{\D}{\D C_{v' l}}\,
    e^{\tr(\mathcal{C}\mathbf{I}_\bullet^t+\mathbf{I}_\circ \mathcal{C}^{-1})}|0\rangle\,,
  \end{aligned}
\end{align}
for $k\neq k'$ and $l\neq l'$, and furthermore
\begin{align}
  \label{eq:vj-action}
  \biggl(\sum_kv_k\,\bar{\mathbf{J}}_{\beta\alpha}^k+\sum_lv_l\,\mathbf{J}_{\beta\alpha}^l\biggr)
  e^{\tr(\mathcal{C}\mathbf{I}_\bullet^t+\mathbf{I}_\circ \mathcal{C}^{-1})}|0\rangle
  =\sum_{k,l}\bar{\mathbf{a}}^k_\alpha\,\bar{\mathbf{a}}^l_\beta \,C_{k l}\,(v_l-v_k)\,
  e^{\tr(\mathcal{C}\mathbf{I}_\bullet^t+\mathbf{I}_\circ \mathcal{C}^{-1})}|0\rangle\,.
\end{align}
Making use of these formulas we can evaluate the action on
\eqref{eq:ansatz-general},
\begin{align}
  \label{eq:m2-eval}
  \!\!\! M_{\alpha\beta}^{(2)}\,|\Psi_{N, K}\rangle
  =\sum_{k,l}\,\biggl(v_l-v_k-p+1-\,\,\,\sum_{\mathclap{i=l-K+1}}^{\mathclap{k+N-K}}\,\,\alpha_i\,\biggr)
  \,\bar{\mathbf{a}}^k_\alpha\,\bar{\mathbf{a}}^l_\beta 
  \int\D\mathcal{C}f(\mathcal{C})\,C_{k l}\,  
  e^{\tr(\mathcal{C}\mathbf{I}_\bullet^t+\mathbf{I}_\circ \mathcal{C}^{-1})}|0\rangle\,.
\end{align}
Here we assumed once more that the boundary terms of the integration
by parts vanish. Furthermore, we used \eqref{eq:diff-ansatz} and
properties of the minors in $f(\mathcal{C})$ similar to
\eqref{eq:deriv-minor}. To ensure Yangian invariance of the ansatz the
parameters $\alpha_i$ have to be chosen such that the bracket in
\eqref{eq:m2-eval} vanishes.

In conclusion, for the ansatz \eqref{eq:ansatz-general} to be Yangian
invariant, the parameters $v_i,c_i$ of the monodromy and the
$\alpha_i$ appearing in this ansatz have to obey the equations
obtained from \eqref{eq:ccbar-eval} and \eqref{eq:m2-eval},
\begin{gather}
  \label{eq:proof-final-ccbar}
  c_k=q\,-\,\,\,\sum_{\mathclap{i=k+1}}^{\mathclap{k+N-K}}\,\,\alpha_i\,,\quad
  c_l=-q\,+\,\,\,\sum_{\mathclap{i=l-K+1}}^l\,\,\alpha_i\,,\quad
  v_k-v_l=-p+1-\,\,\,\sum_{\mathclap{i=l-K+1}}^{\mathclap{k+N-K}}\,\,\alpha_i\,,
\end{gather}
for $k=1,\ldots K$ and $l=K+1,\ldots,N$. These equations are
conveniently addressed after changing from $\{v_i,c_i\}$ to
$\{v^+_i,v^-_i\}$ with \eqref{eq:spec-redef}. In these variables they
are solved by
\begin{align}
  \label{eq:alpha}
  \alpha_i=v^+_{i+K-1}-\,v^-_i+q\,\delta_{i,N-K+1}\,
\end{align}
and imposing the $N$ constraints in
\eqref{eq:spec-perm}. Eq.~\eqref{eq:alpha} turns the ansatz
\eqref{eq:ansatz-general} into the Graßmannian integral formula
\eqref{eq:grass-int}. This concludes the proof of its Yangian
invariance.

\bibliography{literature}{}
\bibliographystyle{utphys}

\end{document}